# Laser Doppler vibrometer and accelerometer for vibrational analysis of the automotive components during Simulink simulation for validation

Fatemeh Rezaei

*Department of Physics, K. N. Toosi University of Technology, 15875-4416, Shariati street, Tehran, Iran*

*Corresponding author*: fatemehrezaei@kntu.ac.ir

**Abstract:** In current research, laser Doppler vibrometer (LDV) as a new diagnostic tool is utilized for non-destructive testing of the automotive equipment. LDV technique is working based on measurement of the Doppler shift of a moving object in an interference set-up. The effects of different noises are considered and eliminated from data analysis. Here, the performance of LDV technique is compared with a reference accelerometer device. Furthermore, a simulation by Matlab Simulink is added to the analysis which confirms the results of the experimental data. Results demonstrated that the laser Doppler vibrometer can measure excellently the frequencies of different automotive components for employing in industry. Therefore, it is proposed that LDV technique can be substituted with other traditional non-destructive testing methods.

## 1. Introduction

Recently, laser Doppler vibrometry (Laser Doppler Velocimetry) has several applications in measuring the velocity of the fluids and their components [1-4], estimating the velocity of moving objects such as the velocity of living organisms, vehicles and aircraft [5, 6], bridge [7], and as well as measuring the speed of the rotating objects such as airplane blades, helicopters [8-11], and also examination of the handling of fruits and vegetables [12-14]. Furthermore, biological measurements such as estimating of the heart rate, respiration rate, blood sugar, blood pressure, and etc.. can be done with this device. It should be mentioned that the ancient methods such as acoustic waves have long been previously utilized for obtaining the velocity and vibration of a moving object [15, 16]. In this way, by examining the signals received from the moving target, the speed and vibration could be acquired. Since, the mentioned methods have disadvantages such as not working properly in unfavorable weather conditions, the need for a material environment to propagate the acoustic waves, and also insufficient accuracy due to the presence of a lot of noise, they are not suitable methods for analysis, but in the proposed method of LDV, in addition to eliminating the above disadvantages, it comprises advantages such as: working in each weather conditions of hot or cold, non-contact measurements, and high speed capability. By using of LDV technique, both of the desired target's vibration and the velocity can be measured simultaneously with just one device and the use of a laser pulse, which means a reduction in costs compared to other conventional methods. In addition, due to the very low divergence of laser beam over long distances and high coherence length of laser pulse, LDV can calculate the vibration and speed of the far moving structures with high accuracy and speed.

Several research groups have measured the speed of moving and rotating car components with the LDV device. For example, Wisler et al [17] measured the velocity of the gas due to the rotation of the compressor rotor by a LDV device. They found that since the growth and separation of the boundary layer was lost due to the rotation of the compressor rotor, the amount of emission generated by the rotor was greater than its actual value. Furthermore, Castellini [18] studied in detail the performance of the Doppler laser vibrometer on various vehicle components, including tires in rotation on the test bench, snow plows, windshield wipers, belts and rubber blades. He studied the temporal and frequency evolutions of the speed and acceleration of these components along with signal processing. In addition, Malikjafarian et al [19] used a Doppler vibrometer for measurement of the damage to the bridge as the vehicle passed over it. They analyzed by simulating 6 vibrometers on a moving vehicle and then, they measured the relative speed between the vehicle and the bridge. They mentioned that the instantaneous curvature of a moving reference caused the curvature of a bridge at an instant in time be sensitive to the local damage. It can be noted that this system was also used for description of the behavior of car window's lifts [20] and car seat belts [21].

In this research, a comparison is performed between the measurements obtained by a constructed laser Doppler vibrometer with a reference accelerometer device (accompanied by amplifier + shaker) on some of automotive elements. Moreover, a simulation is performed by Matlab's simulink for validating the experimental data which affirm the extracted results. It should be mentioned that the techniques employed in this research have been presented to analyze the frequency resonance of some special automotive components which had not been done examined before, according to the literature survey reported.

## 2. Material and Methods: Experimental Analysis

Figure 1 shows a schematic of Michelson's interferometer set-up (Fig.1a) with a packed constructed LDV device (Fig. 1b) for vibrational analysis. The performance of this experimental arrangement is so that the laser light is divided into two parts by a beam splitter. One part is reflected from the fixed mirror without a frequency shift and then arrives to the detector; this beam is the so-called reference beam. The second part of beam hits into the moving target and after that the beam' frequency is shifted by Doppler effect. Then, it is reflected from target and directed into the detector through the beam splitter, this beam is so-called the measurement beam. After the superposition of these two beams on the detector, the interference fringes are formed and with the aid of data acquisition card (DAQ) and signal processing stages, information such as target speed, vibration frequency and displacement will be extracted. The tracking and acquisition data management has been followed by LabVIEW software.

(a)

(b)

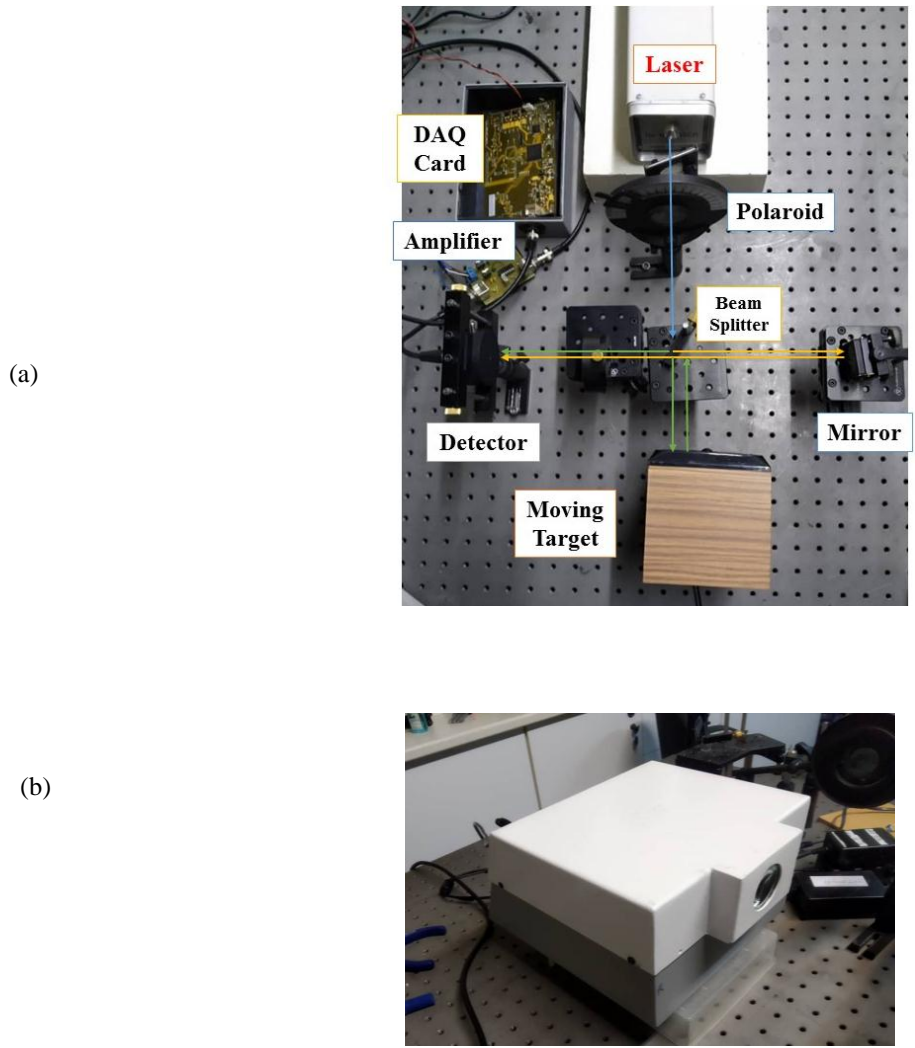

FIGURE 1: A image of (a) an experimental set-up, and (b) constructed and packed laser Doppler vibrometer for measuring the vibration of different moving objects.

In order to perform the experiments, the required programming codes are written with the help of MATLAB and Visual Studio software. The code written in Visual Studio is exploited for communication and data transfer between the detector, data acquisition card and computer. Another code is also written in the Visual Studio media to collect the information with the help of the DAQ card which is shown in the figure 2.

FIGURE 2: Part of the Visual Studio code written for data capturing.

Finally, a signal processing is applied on acquired signal by choosing appropriate software's algorithms and electronic elements. Total noise including of the thermal and Johnson noises, circuit element noises, the noises related to the printer cables, laser shot noise, and as well as background emissions are eliminated by signal processing and appropriate hardware elements. In this research, the noises are removed by moving average algorithm, and as well as employing low, high and band pass filters in different sections of electronic circuits. Simultaneously with LDV device, an accelerometer measures the frequency and displacement of the automotive components as shown in figure 3. It should be mentioned that in accelerometer device, a shaker vibrates the different components of car for providing the arbitrary movements and an amplifier is equipped to this device for enhancement of signal measurement. In this research, different components of car such as pad, air filter and gear are selected for performing the analysis.

**(a)**

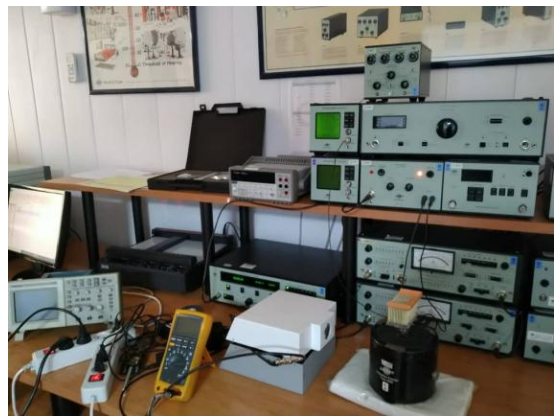

**(b)**

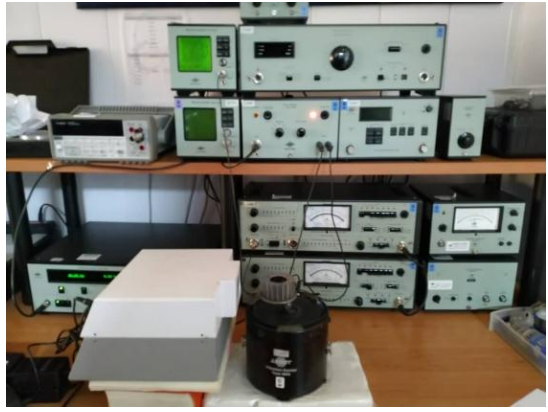

FIGURE 3: A comparison between vibrational measurement with accelerometer and LDV devices on different elements of a) filter, and b) gear.

### 3. Doppler Effect

A specified shift in frequency or wavelength of a wave can be appeared when observer or light source move compared to each other. It should be mentioned that when the light source and the observer are close together, the frequency increases, while when they are far away, the frequency reduces. This difference can be expressed as [11]:

$$f_0 = f(\frac{c}{c \pm v}). \tag{1}$$

here, $f$ is the measured frequency by moving observer, $f_0$ indicates emitted wave frequency, and $c$ and v are light and object velocities, respectively. Since light speed is very larger than observer speed, the above equation can be evolved by Taylor expansion as [11]:

$$f_0 = f(\frac{c}{c \pm v}) \approx f(\frac{c \pm v}{c})^{-1} \approx f(1 \mp \frac{v}{c}) = f + f_d. \tag{2}$$

Generally, Doppler frequency shift is expressed with $f_d = f\,v/c$, but in LDV experiment, the reflected light is calculated. Hence, the Doppler effect must be considered again as [11]:

$$f_d = 2f\frac{v}{c} = 2\frac{v}{\lambda}. \tag{3}$$

here, $\lambda$ is laser wavelength.

### 4. Mathematical Calculations

In Michelson interferometer, the distance between reflector and beam splitter is indicated by $x_r$, and the distance between reference mirror and beam splitter is shown by $x_m$. Here, the related optical phases of the beams in the interferometry system are shown by $\varphi_r = 2kx_r$, and $\varphi(t)_m = 2kx_m$, where, k is wave number $k = 2\pi/\lambda$. In LDV technique, the measurement beam has just time dependent phase, but the reference beam include fixed phase. Moreover, the phase difference between the interference beams is considered as $\varphi(t) = \varphi_r - \varphi_m = 2k\Delta L$, where $\Delta L$ is the vibrational displacement of the target and $\lambda$ is the laser wavelength.

Generally, electrical field in reference beam can be calculated as [22]:

$$E_r = E_{r0} e^{i(\omega_r t + \varphi_r)}. \tag{4}$$

And the measurement electric field can be expressed as [22]:

$$E_m = E_{m0} e^{i(\omega_m t + \varphi_m)} = E_{m0} e^{i((\omega_r + \omega_d)t + \varphi_r - \varphi(t))} = E_{m0} e^{i(\omega_r t + \varphi_r)} e^{i(\omega_d t - \varphi(t))}. \tag{5}$$

The temporal variation of the received intensity by photo detector, at interference point of the reference and measurement is calculated as [22]:

$$I(t) = |E_m + E_r|^2 = I_m + I_r + 2\sqrt{I_m I_r} \cos(\omega_d(t) - \varphi(t)). \tag{6}$$

By considering $K$ as a mixing efficiency coefficient and $R$ as the surface effective reflectivity, intensity equation can be rewritten as [22]:

$$I(t) = I_m I_r R + 2K\sqrt{I_m I_r R} \cos(\omega_d t - \varphi(t)). \tag{7}$$

It should be noted that the phase of intensity presents the displacement, while the frequency indicates the velocity of the moving object.

## 5. Simulation with Matlab Simulink

In this section, a simulation is performed by Matlab Simulink for validating the measurements obtained by LDV method. Generally, LDV is an optical device which detects the Doppler frequency shift which appears when a moving surface scatters the laser beam. This detected frequency shift is directly proportional to the surface vibration and velocity which allows the device to measure the vibration velocity in a non-contact way. It can be stressed that the LDV device includes two interference arms, which is named as the reference and the measurement beams that are described in more detail in the following sections.

### 5.1. Reference Beam

A schematic of reference beam path is shown in Fig. 4. In this figure, the wb and wl blocks show the Bragg-cell frequency shift and the laser frequency, respectively. Here, the × block represents multiplication of the laser frequency by time, and the block Lr indicates the loss generated by the optical elements of beam lenses and splitters. In addition, the sin block denotes the sine of the total frequency multiplied by the time, for the reference beam signal.

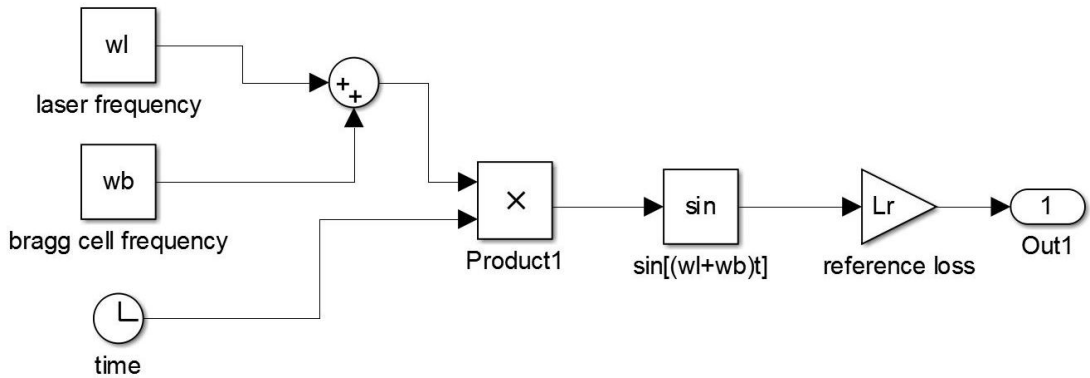

FIGURE 4: The reference beam's arm in the Simulink calculation.

## 5.2. Measurement beam

A schematic of operation of the LDV device in the measurement beam is displayed in Fig. 5. In this figure, the velocity block represents the velocity of moving object at different time. Furthermore, the sin block indicates the sin of laser frequency plus Doppler frequency shift. Similar to the reference beam, Lm block represents the noises produced by the measurement of the optical elements.

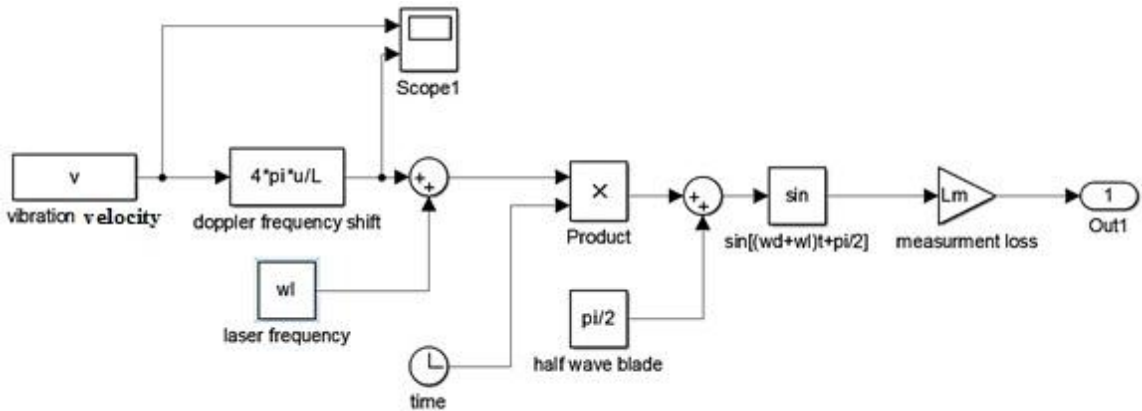

FIGURE 5: The measurement beam's path in LDV system.

## 5.3. Noises in LDV technique

In all measurement diagnostic techniques, the accuracy of estimation is limited by system noises [23, 24]. It is worth nothing that in current research, the noises can generate some errors in measurements precision of LDV data which include: i) Thermal noise, ii) Shot noise, iii) Speckle noise, iv) and Flicker noise.

Each of these mentioned noises are explained in detail in following sections, however, most of time, the magnitude of thermal and shot noises overcome the amount of flicker and speckle noises. It should be mentioned that in order to increase the reliability of the measurements, all constant values for these noises are inserted according to the experimental condition of the LDV set-up.

### 5.3.1. Shot noise

In LDV device, shot noise is essentially generated by electronic circuits and laser source, such as photo detector (PD), data acquisition card (DAQ) and other controller circuits. Due to photons' population fluctuations, this noise can be appeared in PD output signals which is calculated as follows [25, 26]:

$$i_{shot} = \sqrt{2ei_{signal}\Delta f}, \tag{8}$$

where, $i_{signal}$, $e$ and $\Delta f$ are the signal current, electron charge and signal bandwidth, respectively.

### 5.3.2. Thermal noise

Because of the random motion of electrons in a resistant and conductive material, thermal (or Johnson-Nyquist) noise is generated. Thermal noise current can be calculated by knowing the total resistance of detector and its amplifier in the system terminal voltage as [23]:

$$i_{thermal} = \sqrt{\frac{4K_B T \Delta f}{R_L}} = \sqrt{4K_B T \Delta f \frac{R_i + R_d}{R_i R_d}}. \tag{9}$$

here, $T$, $K_B$, $R_L$, and $\Delta f$ indicate absolute temperature, Boltzmann constant, total resistance, and signal bandwidth, respectively. Moreover, $R_i$ and $R_d$ are the input resistance of the amplifier and the detector. It should be noted that in this research, the effective resistance value is considered as 50 ohms.

### 5.3.3. Flicker noise

Flicker noise is the dominant noise in the low frequencies which is generated in all of the semiconductors because of the oscillations of silicon resistance and crystal defects [27]. This noise is calculated from the below equation as [28]:

$$i_{flicker} = \sqrt{K \frac{i_{signal}^{\alpha}}{f^{\beta}} \Delta f}. \tag{10}$$

where, $K$, $\alpha$ and $\beta$ are constant parameters dependent to processing of material. $\Delta f$, $f$ and $i_{signal}$ are the system bandwidth, signal frequency and signal current, respectively.

### 5.3.4. Speckle noise

Speckle noise depends on surface quality of moving object and is created due to interference of several waves with the same frequencies. Power function can be computed from intensity distribution as follows [29]:

$$P(I)dI = \frac{1}{\bar{I}}\exp(\frac{I}{\bar{I}})dI, \tag{11}$$

here, $I$ and $\bar{I}$ are the intensities of the laser beam and its mean value, respectively.

### 5.4. Signal interference including both of reference and measurement beams

As clearly shown in Fig. 6, the two references and measurement beams are collided together in a photodetector. Since the second power of the field or the intensity value for the detector is meaningful, therefore the calculated parameter in the corresponding block is powered to two. It should be mentioned that in this simulation, the related signals of each measuring and reference beam can be individually observed in the oscilloscope.

Finally, the influences of all the noise are considered and they are subtracted from the interference output signal to obtain the exact and net result of the output vibrating signal.

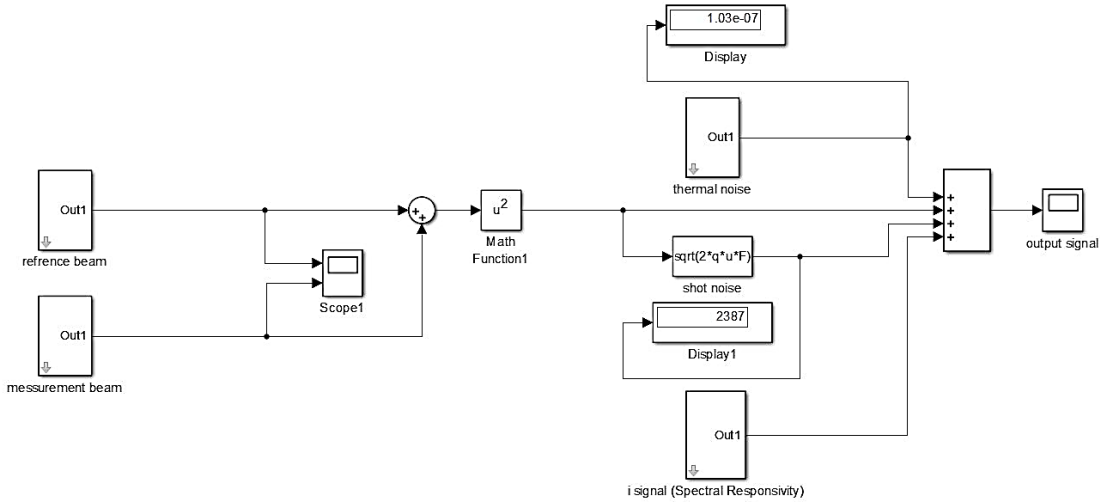

FIGURE 6: The final block diagrams of the Simulink calculation by considering all effective noises.

It should be noted that the frequency of the vibration can be calculated by considering the fast Fourier transform of the temporal signal by insertion of FFT block. Since, due to Eq. (7), a cos function is appeared, therefore the FFT of sin and cos functions must be considered as follows:

$$FFT(\sin 2\pi at) = \int_{-\infty}^{+\infty} e^{-i2\pi\upsilon t} \sin 2\pi at \, dt = \frac{1}{2i}[\delta(\upsilon - a) - \delta(\upsilon + a)]. \tag{12}$$

$$FFT(\cos 2\pi at) = \int_{-\infty}^{+\infty} e^{-i2\pi\upsilon t} \cos 2\pi at \, dt = \frac{1}{2}[\delta(\upsilon - a) - \delta(\upsilon + a)]. \tag{13}$$

## 6. Results and discussion

Automotive manufacturers and suppliers are following to enhance the product quality according to the costumer's demands by applying different non-destructive testing method. In general, frequency analysis of time history helps to find several natural frequencies of the automotive components for evaluation of their efficiencies. On the other hand, the interpretation of the meaning of each peak indicates the tracking performance of each element. Here, after an introducing and discussion of the measurement technique, concentration is focused on the various applications in the measurement of vibration of some typical automotive component. In current study, the measurements are done during out-of-plane vibrations of automotive components by two devices of laser Doppler vibrometer and accelerometer. In order to present the accuracy of these methodologies, some experimental measurements are performed on different mechanical car components. For instance, here first, in order to evaluate the failures and quality performance of the pinion gear, it is excited with an impact shaker. All response signals are guided into a FFT analyzer, which consists of a data acquisition card and amplifier boards and signal processing software and hardware for further analysis. In this research, the vibration signals are collected by photodetector and stored in the computer. It should be mentioned that the sampling frequency must be sufficient enough to record the approximately frequency content of vibration. Here, the acquired signals have relatively low frequencies in order of Hz, therefore, the sample rate of 20 KHz is chosen for the whole of experiment.

In this stage, in order to present the accuracy of this research, some experimental measurements are performed by LDV device. First of all, testing is performed on pinion gear due to its vibrational signal is significant in detection and diagnosis of gearbox damages and faults. The most common gear wear mechanisms, such as scouring, micropitting, and pitting frequently are considered as the failures which directly influence on the response frequency. The different components in car which are placed in combination with pinion gear are shown in figure 7.

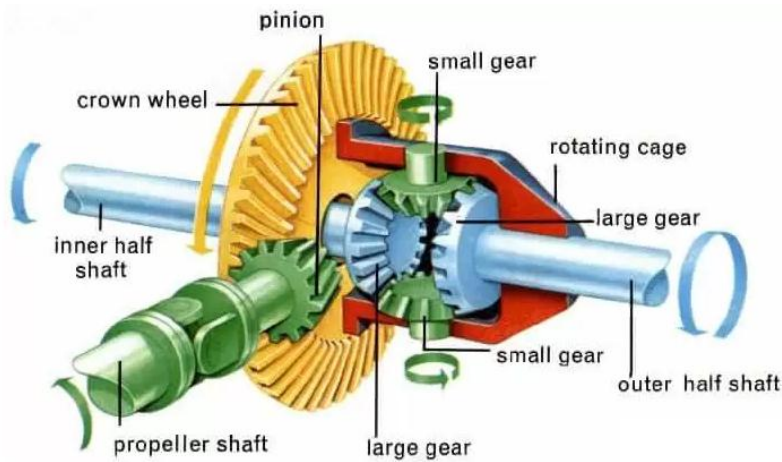

FIGURE 7: Different components related to the pinion gear.

Exploration on the gear damage utilizing the vibration signal is important. Since, gears vibration signals are difficult and complex to analyze. The spectra obtained from the analysis of two pinion gears by LDV technique is plotted in the figure 8. Here, indeed the noise generated by different factors of optics and electronics are eliminated and then, the vibrational signal is presented. Here, two frequencies of 85 and 76 Hz are measured for two safe pinion gears which represent that the other gears related to chosen company including frequencies in another interval must encounter to some different failures such as cracks, deformation, and etc.. In addition, a little difference in frequency response in figure 8 may be attributed to alteration in the stiffness of the gear tooth or the constituent elements.

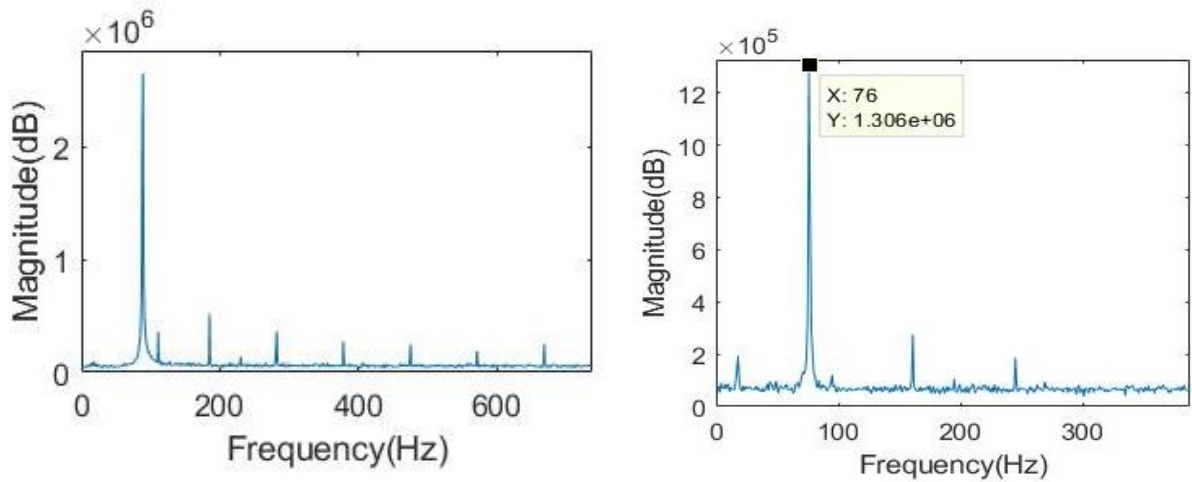

FIGURE 8: Frequency evolution of the pinion gear vibrations with a frequencies of (a) 85 Hz, and (b)76 Hz, using a Doppler laser vibrometer.

It should be mentioned that these results are in approximately good agreement with other literatures which represents the high accuracy of this method. During comparison with other references, a broad range of numbers was reported for response frequencies of the automotive pinion gear depending on size, gender, and constituent materials. For instance, Ref. [30] represented a solution for the motion equation by different methods of the perturbation method, the Ritz method, the discretization method, and the stepwise time integration for estimation of non-linear torsional vibration of a one-stage transmission gear system. They reported the angular frequency response of the system in the ranges of 0.6-1.4 rad/s. Furthermore, Ref. [31] evaluated the torsional displacement of the pinion along the x and y axis in order of 13 to 222 Hz during theoretically and experimentally analysis. In addition, Ref. [32] explained the frequency response function of gear and frame system for various fault severities in order of 1 KHz to 4 KHz during experimental analysis of stiffness accompanied by modal testing in conjunction with a theoretical model. Consequently, the diversities in the reported values for pinion gear can be attributed to different elements utilized in various factories.

It should be noted that for the geared system operating, different modal tests can help to find directly the severity of the damaged components as a failure diagnosis. In current research, based on Fourier analysis of the vibration produced by the shaker as an innovative way, the performance of various automotive components is evaluated.

Noticed that that due to the non-contact feature of the measurement, no influence of mechanical damage needed to be taken into account. This enormously increases the data quality for the performing experiment far from environmental harm. Then, the vibration amplitude of the safe pad component is shown in figure 9. As it is clearly seen in this figure, a peak at 6 Hz presents the natural frequency of a chosen automotive component of pad. It is worth noting that natural frequency is a critical point which impact on the safe or damage behavior of the mechanical components.

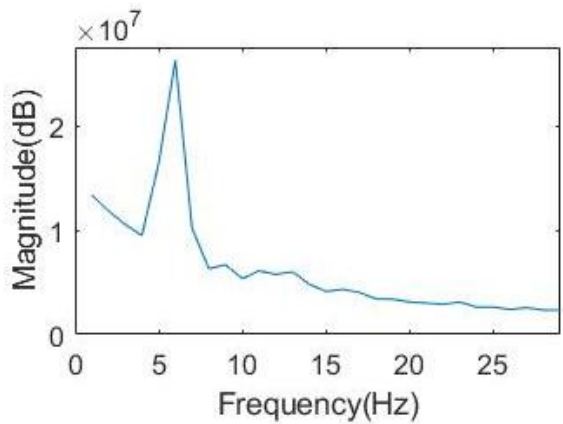

FIGURE 9: Frequency distributions of the pad vibrations with a peak at 6 Hz estimated by using of a Doppler laser vibrometer.

In order to estimate the vibrational behavior of the air filter under shaking operative condition and to check the performance of LDV device, the measurements are represented in figure 10. Figure 10 illustrates the experimentally measured frequency during vibration of sample when the noises are omitted. Here, the vibration signals generated by the tested air filter are monitored for different points and the average signal is reported.

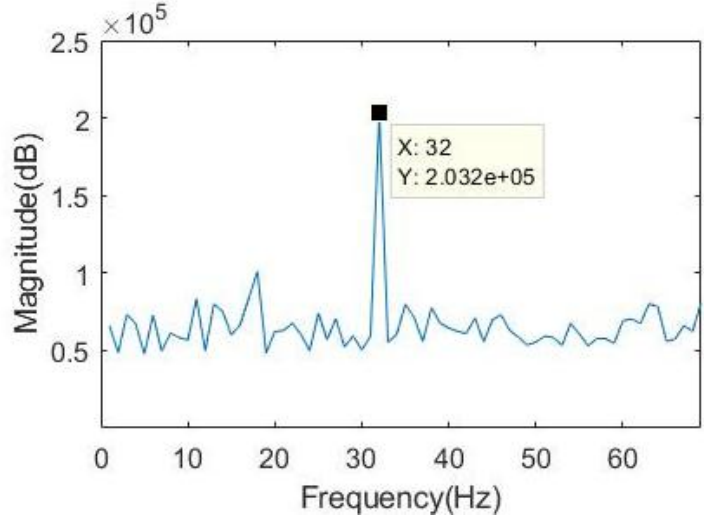

FIGURE 10: Frequency variations of the air filter vibrations including a peak appeared at 32 Hz measured by using a Doppler laser vibrometer.

During comparison with other researches, it is found that little groups studied the vibrations of air filter. For instance, Beigmoradi et al in Ref. [33] performed the modal analysis of a concept filter box and presented some modifications for increasing the vibration performance of introduced system. They employed finite element method (FEM) for estimation of the frequency response function (FRF) and reported frequencies between 87 Hz to 350 Hz. It should be mentioned that this group investigated on filter box, but not the sole filter. Therefore, a logical comparison is not possible in this case.

Furthermore, in order to understand time-dependent frequency characteristics in signals, vibration data must be further processed by noise reduction. It should be noted that for all the cases examined, different hardware and software methodologies are employed for noise elimination such as Henderson moving average method and utilizing of different low, high and band pass filters.
From the frequency evolutions in figures 8, 9 and 10, it can be concluded that the elements in other ranges must be encountered with different harms or damages such as internal cracks [34,35]. Then, simultaneously a measurement is performed by two devices of LDV and accelerometer for evaluating the accuracy of the constructed LDV which is shown in Fig. 11. In order to verify the efficiency of an accelerometer, it has been mounted on the different automotive components. In figure 11, the accelerometer measured 40 Hz for vibrations of a typical air filter, while LDV calculates 39.9 Hz which shows relatively similar frequency magnitudes with a low error value. Then, a repetition on measurements is performed on different automotive elements containing various frequencies, so that laser Doppler vibrometer and accelerometer frequency histories presented relatively good accommodation. Finally, the comparison results between the measurements on different car components by LDV technique and accelerometer device are summarized in table 1 which shows low error values. Moreover, the amplitudes of vibrations are illustrated in this table. As it is obviously seen, the maximum value for error is about 0.30 which can be considered as a low magnitude.

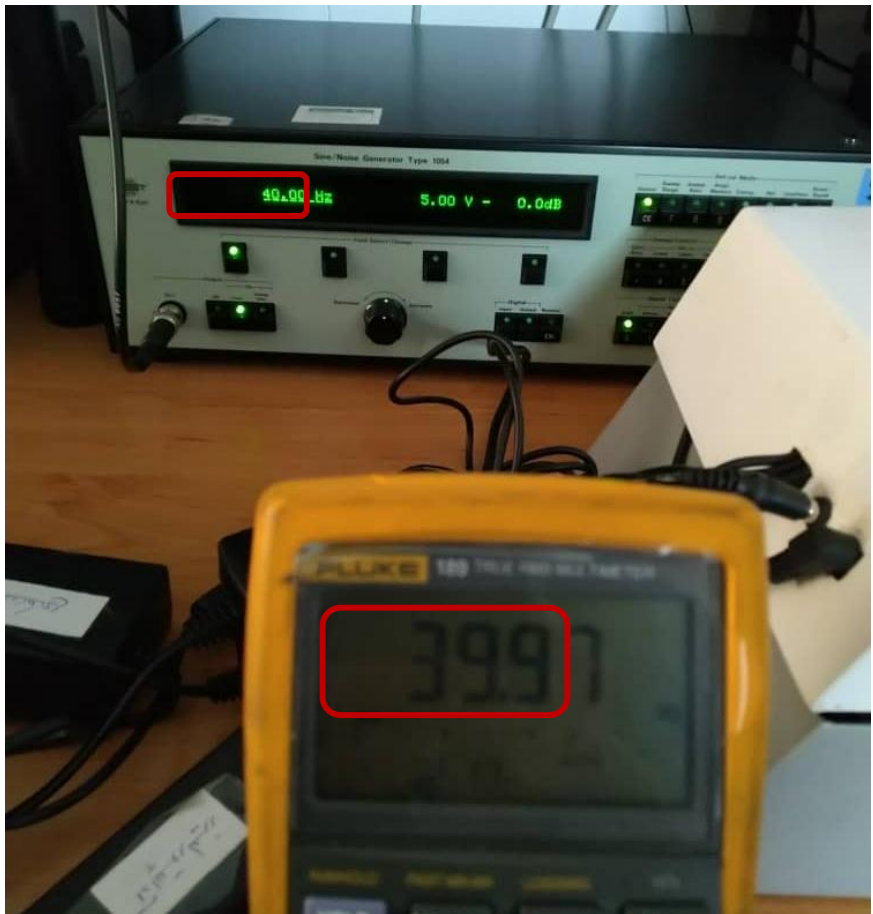

FIGURE 11: A comparison between frequency measured by LDV technique and accelerometer device which illustrate relatively a good agreement.

TABLE 1: Results of the comparison between frequency measured by laser Doppler vibrometer and an accelerometer device. Here, calibration data are measured by accelerometer, and <U.U.T> indicates the results with LDV device.

| Test point | <Calibrator> Applied Value | | <U.U.T> Indicated Value |
|---|---|---|---|
| | Frequency (Hz) | Displacement | Frequency (Hz) |
| 1 | 10.000 | 163 µm | 10.00 ± 0.30 |
| 2 | 20.000 | 4.15 mm | 20.00 ± 0.20 |
| 3 | 30.000 | 1.68 mm | 30.00 ± 0.15 |
| 4 | 40.000 | 336 µm | 40.00 ± 0.15 |
| 5 | 50.000 | 184 µm | 50.00 ± 0.15 |
| 6 | 60.000 | 1.70 mm | 60.00 ± 0.15 |
| 7 | 70.000 | 434 µm | 70.00 ± 0.15 |
| 8 | 80.000 | 1.16 mm | 80.00 ± 0.15 |
| 9 | 90.000 | 165 µm | 90.00 ± 0.15 |
| 10 | 100.000 | 770 µm | 100.00 ± 0.15 |
| 11 | 130.000 | 401 µm | 130.00 ± 0.15 |

Finally, the results obtained from Simulink simulation is represented in Fig. 12. In this step, by observation of the Simulink calculation results, complete and acceptable conformity can be attained between the input data and the prediction with the Simulink simulation. Furthermore, as it is clearly seen in this figure, the simulation results are in relatively good accordance with the experimental analysis in Figs. 9 and 10 with an acceptable error, thus indicating the correctness of the LDV experimental results. On the other hand, the ability of the LDV technique in response frequency analysis can be proved in this study. It should be noted that the differences in the order of heights in these figures are due to the fact that they are calculating different parameters which are not the purpose of this research and only the frequency matching is the main goal.

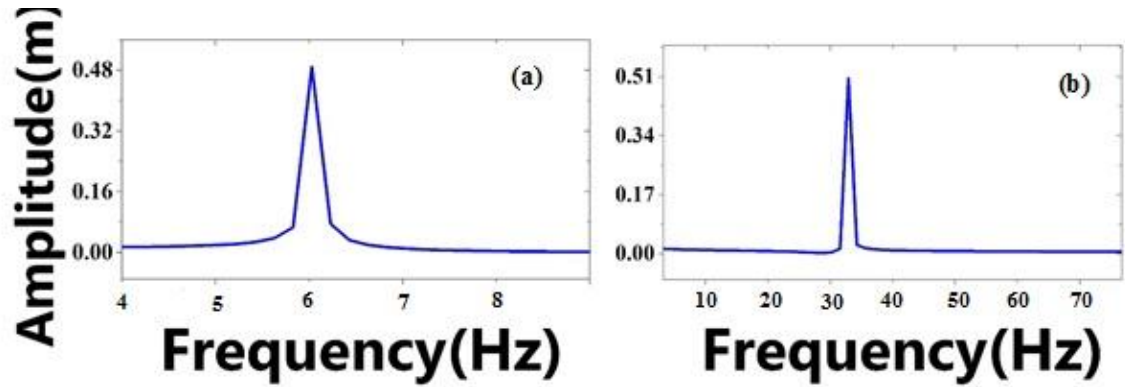

FIGURE 12: Vibration calculation by Matlab Simulink for (a) pad, and (b) air filter which is in consistent with the results of Figs. 9 and 10.

In Fig. 12, as expected according to Eqs. (12) and (13), a Dirac delta function presents in all of the figures, due to appearance of a constant value for the vibrational frequency.

In the next step, a comprehensive comparison between the data measurements by LDV method, accelerometer device, and Simulink simulation is shown in table 2. According to the results presented in this table, it can be clearly realized that there is a little difference among the frequencies predicted by LDV technique, accelerometer device, and Simulink results or a close agreement between simulation and experimental values are viewed. This observed variance can be attributed to the difference of noises nature in two used devices which produce some measurement errors.

TABLE 2: A comparison between frequencies measured with LDV, accelerometer and Simulink calculations.

| Automotive Components | Frequency (Hz) by LDV | Frequency (Hz) by accelerometer | Frequency (Hz) by simulink |
|---|---|---|---|
| pad | 6 | 5.9 | 6.1 |
| air filter | 32 | 31.3 | 33 |
| gear | 76 | 75.5 | 77 |

## 7. Conclusion

In current research, the performance of different components of automotive such as pads, air filters and gears were evaluated by two devices of laser Doppler vibrometer and a accelerometer for verification of the performance of these devices. Moreover, for more illustration, a simulation was done with Matlab Simulink for confirming the experimental analysis. The good agreement between the simulation results and the measured ones illustrated the possibility of effectively reduction of the noise contributions in vibrational frequency evaluation. The results illustrated that the LDV method can be substituted by old mechanical techniques for non-destructive testing. Furthermore, the proposed measurement systems and the results represented can be useful to better understand the dynamic behavior of automotive elements under operative circumstances.

**Author Contributions Statement**
Fatemeh Rezaei performed whole sections of this manuscript.

**Competing interests**
I declare that the author has no competing interests.

**Funding**
This work was supported by the Iran National Science Foundation (INSF) [97007912].

**Data Availability Statement**
The data used to support the findings of this study are available from the corresponding author upon request.

**Acknowledgment**
The author is thankful to Iran National Science Foundation (INSF) for the given individual research chair and calibration center of Pishtazan Industrial Development Company of Aria Razi for confirming the performance of LDV device.